\newcommand{\bee}{\begin{equation}}
\newcommand{\ene}{\end{equation}}
\newcommand{\lt}{\left}
\newcommand{\rt}{\right}

\documentclass[journal]{IEEEtran}

\ifCLASSINFOpdf
\else
\fi
\usepackage[cmex10]{amsmath}
\usepackage{amssymb}
\usepackage{graphicx}

\begin{document}

\title{Shielding of a small charged particle in weakly ionized plasmas}

\author{Manis Chaudhuri, Sergey A. Khrapak, Roman Kompaneets, and Gregor E. Morfill 
\thanks{This work was supported by DLR under Grant 50WP0203. R. Kompaneets is the recipient of a Professor Harry Messel Research Fellowship
supported by the Science Foundation for Physics within the University of Sydney.}
\thanks{M. Chaudhuri, S. A. Khrapak, and G. E. Morfill are with
Max-Planck-Institut f\"ur extraterrestrische Physik,
D-85741 Garching, Germany (email: chaudhuri@mpe.mpg.de;
skhrapak@mpe.mpg.de; gem@mpe.mpg.de). R.~Kompaneets is with the School of Physics,
The University of Sydney, NSW 2006, Australia.}}




\maketitle

\begin{abstract}
In this paper we present a concise overview of our recent results concerning the electric potential distribution around
a small charged particle in weakly ionized plasmas. A number of different effects which influence plasma screening properties are considered. Some consequences of the results are discussed, mostly in the context of complex (dusty) plasmas.
\end{abstract}


\IEEEpeerreviewmaketitle

\section{Introduction}

Complex (dusty) plasmas are plasmas containing small charged
particles of solid matter (dust grains). These particles are usually large enough to be observed
individually, which allows experimental investigation with high
temporal and spatial resolution. Hence, complex plasma is a valuable model system for studying various phenomena (e.g, phase
transitions, self-organizations, waves, transport, etc.) at the most
elementary kinetic level~\cite{Morfill_Springer,ShuklaIOP,Fortov2005PR,Fortov2004UFN,Vladimirov2004PR}.

The character of the plasma-particle as well as the inter-particle interactions appears to be one of the most fundamental
questions for understanding the physics behind the observed phenomena in laboratory
complex plasmas~\cite{Morfill_Springer,ShuklaIOP,Fortov2005PR,Fortov2004UFN,Vladimirov2004PR} as well as in astrophysical plasmas~\cite{Mendis}, plasma of fusion
devices~\cite{Winter,Pigarov,Angelis}, plasma processing~\cite{Vladimirov2004PR,Bouchule,Kersten}, etc. Of particular importance are basic processes such as particle charging, electric potential distribution around a charged particle in plasmas, interparticle interactions, momentum and energy transfer between different complex plasma components etc.~\cite{KhrapakCPP}.

The main aim of this paper is to present a brief overview of our recent results (mostly
theoretical) related to the electric potential distribution around a charged test particle
in plasmas. A number of different effects which influence plasma screening properties are considered, such as plasma absorption on the particle surface, non-linearity of ion-particle interaction, ion-neutral collisions, plasma production and loss processes in the vicinity of the particle. These effects influence and often completely determine the shape of the electric potential around the particle, especially its long-range asymptote. This is important for a number of collective properties of complex plasmas (e.g., interparticle coupling, phase transitions, phase diagrams, transport, etc). Below, we first discuss the role of these effects in isotropic plasma conditions (Section \ref{iso}) and then consider the case of anisotropic plasmas (Section \ref{aniso}). This is followed by a short summary in Section \ref{concl}.

\section{Isotropic plasmas}\label{iso}

The distribution of the electric potential around a small individual spherical {\it non-absorbing} particle of radius $a$ and charge $Q$ in isotropic plasmas is often described by the Debye-H\"{u}ckel
(Yukawa) form
\begin{equation} \label{Yukawa}
\phi(r)\simeq\phi_{\rm s}(a/r)\exp\left[-(r-a)/\lambda_{\rm D}\right]\simeq
(Q/r)\exp(-r/\lambda_{\rm D}),
\end{equation}
where $\lambda_{\rm D}$ is the linearized Debye length, $\lambda_{\rm D}=\lambda_{{\rm D}i}/\sqrt{1+(\lambda_{{\rm D}i}/\lambda_{{\rm D}e})^2}$. The ion (electron) Debye length is defined as $\lambda_{{\rm D}i(e)}=\sqrt{T_{i(e)}/4\pi e^2n_{i(e)}}$, where $T_{i(e)}$ is the ion (electron) temperature, $n_i\simeq n_e\simeq n$ is the unperturbed plasma density, and $e$ is the elementary charge. The particle surface potential $\phi_{\rm s} = (Q/a)\exp(-a/\lambda_{\rm D})$ is usually of the order of the electron temperature,  $\phi_{\rm s}\sim -T_e/e$, due to much higher electron mobility. The above expression can be obtained by solving the linearized Poisson equation with the assumption that ion and electron densities
follow Boltzmann distributions and the condition $|e\phi_{\rm s} / T_{i(e)}| \lesssim 1$ is satisfied. It is to be noted that usually in complex plasmas  $T_e\gg T_i$ and the linearization is often invalid since ion-particle coupling is very strong close to the particle. Nevertheless, numerical solution of
the non-linear Poisson-Boltzmann equation shows that the functional form of
Eq.~(\ref{Yukawa}) still persists, but the actual value of the particle charge should be
replaced by an effective charge which is somewhat smaller than the actual one in the absolute magnitude~\cite{Nefedov1998, Bystrenko-Zagorodny1999}.

The effect of plasma absorption on the particle surface strongly modifies
the potential distribution. The continuous ion and
electron fluxes from the bulk plasma to the particle make their
distribution functions anisotropic in the velocity space. Although the deviations are negligible
for repelled electrons~\cite{Alpert}, for
attracted ions they are quite substantial. In the absence of plasma
production and loss in the vicinity of the particle, conservation of
the plasma flux completely determines the long-range asymptote of the
potential and in  {\it collisionless plasmas} it scales as $\phi(r)\propto
r^{-2}$~\cite{Fortov2005PR,Alpert,Tsytovich1997}. Close to the particle (up to a distance of few Debye
radii from its surface), the Debye-H\"{u}ckel (DH) form works
reasonably well. However, in the regime of strong ion-grain coupling, the linearized Debye length $\lambda_{\rm D}$ should be replaced by the effective screening length $\lambda_{\rm eff}$.
The exact dependence of $\lambda_{\rm eff}$ on plasma and grain parameters is not known for the general case and so far it was determined using numerical simulations only for a limited number of special cases \cite{Daugherty,LampePoP2003,RatynskaiaPoP2006}. For a collisionless plasma with $T_e\gg T_i$ a fit based on numerical results of Ref.~\cite{Daugherty} has been recently proposed \cite{KhrapakCPP}. The corresponding expression is $\lambda_{\rm eff}\simeq\lambda_{\rm D}[1+0.105\sqrt{\beta}+0.013\beta]$, where $\beta=|Q|e/(T_i\lambda_{\rm D})$ is the so-called scattering parameter \cite{KhrapakPRE2002}, which is a natural measure of nonlinearity in ion-grain interaction. This fit has been used to calculate the ion drag force in a collisionless plasma with strong ion-grain coupling in Ref.~\cite{KhrapakPoP2009} and to estimate the effect of polarization interaction on the propagation of dust acoustic waves in complex plasmas in Ref.~\cite{KhrapakPRL2009}.

Another important factor which influences the structure of the
electric potential around an absorbing particle in plasmas is
ion-neutral collisions. In the weakly collisional regime ($\ell_i\gtrsim\lambda_{\rm D}$, where $\ell_i$ is the ion mean free path) the infrequent
ion-neutral collisions create trapped ions which increase the ion density in
the vicinity of the particle. This affects the ion flux collected by the particle~\cite{Lampe2001,Ratynskaia2004PRL,Khrapak2005PRE} and, therefore, modifies the potential distribution.
In strongly collisional plasmas ($\ell_i \ll \lambda_{\rm D}$) the electric potential is known to exhibit a Coulomb-like ($\propto r^{-1}$) decay \cite{Su-Lam1963,Bystrenko-Zagorodny2003,Khrapak2006b,Filippov2007a}. A transition from the DH to the unscreened Coulomb potential with increasing ion
collisionality was observed in a numerical simulation~\cite{Zobnin}.

Recently, a simple linear kinetic model has been proposed independently by Filippov
{\it et al}.~\cite{Filippov2007b} and Khrapak {\it et al}.~\cite{Khrapak2008} which takes into account the combined effect of ion absorption on the particle and ion-neutral collisions. Using this model the electric potential distribution can be calculated in the entire range of ion collisionality. Below we present the general expression for the potential obtained in Ref.~\cite{Khrapak2008} and analyze some interesting limiting cases.

\subsection{Linear kinetic model}

In this model a small (point-like) negatively charged individual grain immersed in a stationary isotropic weakly
ionized plasma is considered. Plasma production and loss processes are neglected in the vicinity of the particle
except on the particle surface, which is fully absorbing. The electron density satisfies the Boltzmann relation, and ions are described by the kinetic equation accounting for ion-neutral collisions and ion loss on the particle surface. Collisions are modeled by the Bhtanagar-Gross-Krook (BGK) collision integral with a velocity independent (constant) effective ion-neutral collision frequency $\nu$. Ion loss is expressed through an effective (velocity dependent) collection cross section $\sigma$. The linear response formalism is used to solve the Poisson equation along with the corresponding equations for ions and electrons. The resulting expression for the electric potential is~\cite{Khrapak2008}
\begin{equation}\label{potential_Khrapak}
\phi(r)=\frac{Q}{r}\exp(-k_{\rm D}r)-\frac{e}{r}\int_0^{\infty}\frac{k_{\rm
D}\sin(kr)f(\theta)dk}{k^2+k_{\rm D}^2}\equiv\phi_{\rm I}+\phi_{\rm II},
\end{equation}
where
\begin{displaymath}
f(\theta)=\frac{8n}{\pi^{3/2}k_{\rm
D}}\frac{\int_0^{\infty}\sigma(\zeta)\zeta^2\arctan(\zeta/\theta)\exp(-\zeta^2)d\zeta}{1-\sqrt{\pi}\theta\exp(\theta^2)[1-{\rm
erf}(\theta)]}.
\end{displaymath}
Here $k_{{\rm D}i(e)}=\lambda_{{\rm D}i(e)}^{-1}$ is the inverse ion (electron) Debye length, $k_{\rm D}=\sqrt{k_{{\rm D}e}^2 + k_{{\rm D}i}^2}$, $v_{T_i}=\sqrt{T_i/m_i}$ is the ion thermal velocity, $m_i$ is the ion mass, $\theta=(\nu/\sqrt 2 kv_{T_i})$,
and $\zeta^2=v^2/2v_{T_i}^2$. The first term $\phi_{\rm I}$ in Eq.~(\ref{potential_Khrapak}) is
the familiar Debye-H\"{u}ckel potential. The second term $\phi_{\rm II}$ appears due to
ion absorption by the particle and accounts for ion-neutral collisions. For a
non-absorbing particle [$\sigma\equiv 0$] only the conventional DH form survives, as
expected. In this case ion-neutral collisions do not affect the potential distribution.

Let us now consider some important limiting cases.

\subsubsection{Collisionless limit}

In the collisionless\index{collisionless regime} (CL) limit one can use
the OML collection cross section $\sigma(\zeta)=\pi
a^2\left[1+(z\tau)\zeta^{-2}\right]$ to get~\cite{Filippov2007a}
\begin{equation}\label{potential1}
\phi_{\rm II}(r)=-\frac{e}{r}\frac{\pi a^2 n (1+2z\tau)}{2k_{\rm D}} {\mathcal
F}(k_{\rm D}r),
\end{equation}
where ${\mathcal F}(x)=\left[{\rm e}^{-x}{\rm Ei}(x)-{\rm e}^x{\rm Ei}(-x)\right]$ and
${\rm Ei}(x)$ is the exponential integral. Here, $z=|Q|e/aT_e$ is the normalized particle charge and $\tau = T_e/T_i$ is the electron-to-ion temperature ratio. For sufficiently large distances $x\gg 1$, ${\mathcal F}(x)\approx
2/x$ and the corresponding potential is
\begin{equation}\label{asymptote_CL}
\phi_{\rm II}(r)\simeq-\frac{T_e}{e}\left(
\frac{a}{r}\right)^2\frac{1+2z\tau}{4(1+\tau)},
\end{equation}
which coincides with the well known result of probe theory~\cite{Fortov2005PR,Alpert,Tsytovich1997,Allen2000}. The long-range asymptote of the potential scales as $\propto r^{-2}$ in this collisionless limit.

\subsubsection{Strongly collisional limit}

In the opposite strongly collisional (SC) regime the actual
form of $\sigma(\zeta)$ is not important since the integral in
$f(\theta)$ is directly expressed through the ion flux $J_i$ in this
case. The resulting potential is
\begin{equation}\label{potential2}
\phi_{\rm II}(r)\simeq-\frac{e}{r}\frac{J_i}{D_ik_{\rm D}^2}\left[1-\exp(-k_{\rm D}r)\right],
\end{equation}
where $D_i=v_{T_i}^2/\nu$ is the diffusion coefficient of the ions. This expression
coincides with the results obtained using the hydrodynamic approach~\cite{Khrapak2007a,Khrapak2007b}.
The long-range asymptote of the potential decays as $\propto r^{-1}$ in this case.

\begin{figure}[!t]
\centering
\includegraphics[width = \linewidth]{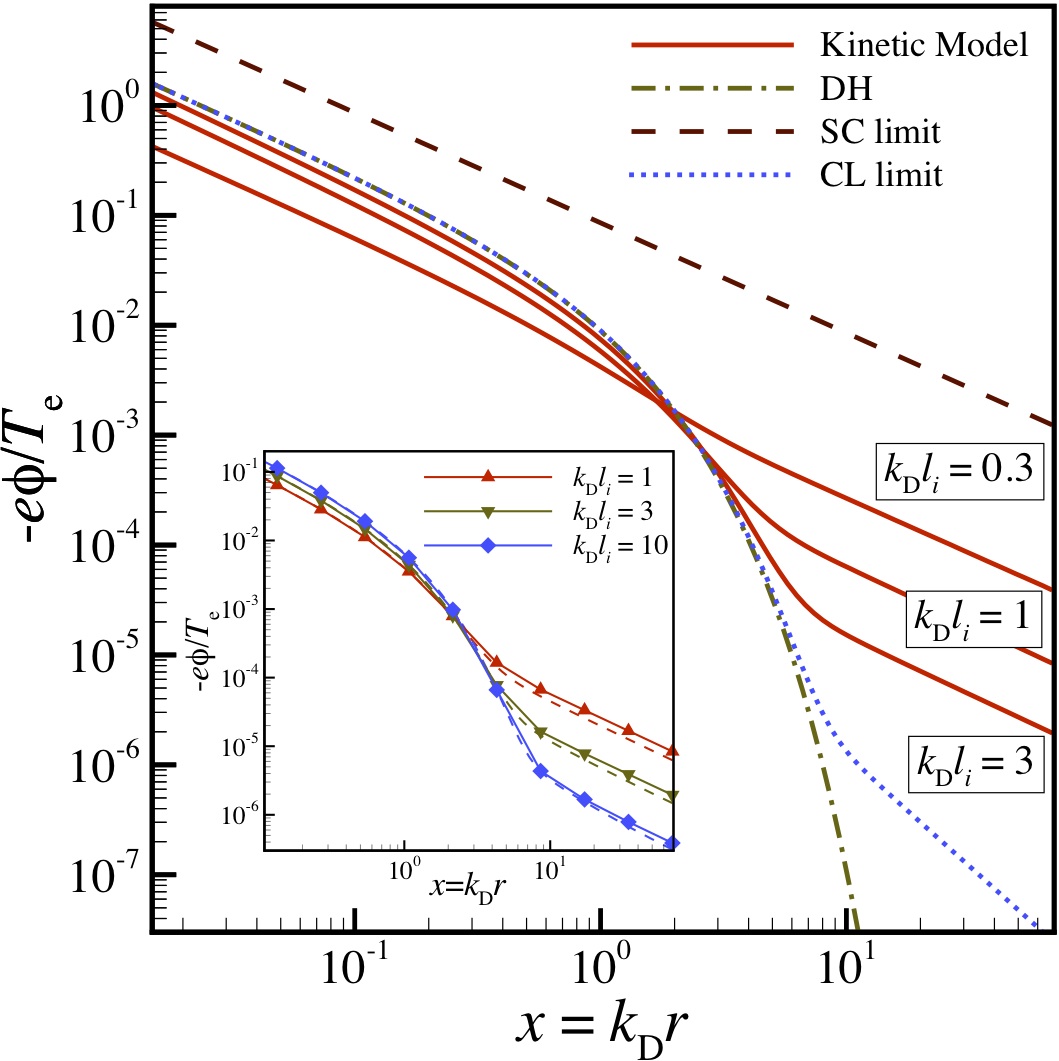}
\caption{Distribution of the normalized electric potential around a small individual charged particle in an isotropic weakly ionized plasma for different values of the ion collisionality index.
The solid curves are obtained using numerical integration of Eq.~(\ref{potential_Khrapak}).
Dashed (dotted) curve corresponds to the analytic approximation in the strongly collisional (collisionless)
limit. The dash-dotted curve shows the Debye-H\"{u}ckel potential with the surface potential calculated
from the (collisionless) OML model. The inset shows a comparison between direct numerical integration of Eq.~\ref{potential_Khrapak} (solid lines with symbols) and analytical
approximation for the weakly collisional regime [Eqs. (\ref{potential_Khrapak}) and (\ref{potential3})].}\label{f1}
\end{figure}

\subsubsection{Weakly collisional limit}

The most interesting regime relevant to many complex plasma
experiments in gas discharges is the weakly collisional (WC)
regime\index{weakly collisional regime}, $\ell_i \gtrsim \lambda_{\rm
D}$. In this case the functional form of $\sigma(\zeta)$ is required to
calculate the potential distribution. A simple assumption of a constant cross section $\sigma (\zeta)= \sigma_0=
\sqrt{\pi/8}(J_i/nv_{T_i})$ was made
in Ref.~\cite{Khrapak2008}, which allowed to avoid divergence of the integrals in calculating
$f(\theta)$. The result is
\begin{multline}\label{potential3}
\phi_{\rm II}(r)\simeq-\frac{e}{r}\frac{\sqrt{\pi}}{4\sqrt{2}}\frac{J_i}{k_{\rm
D}v_{T_i}} \\ \times \left\{{\mathcal F}(k_{\rm D}r)+\frac{3.34}{\ell_i k_{\rm
D}}\left[1-\exp(-k_{\rm D}r)\right] \right\}.
\end{multline}
The two terms in the curly brackets of Eq.~(\ref{potential3}) correspond to absorption induced
``collisionless'' and ``collisional'' contributions, respectively. The collisional
contribution to the potential dominates over the collisionless one for $r\gtrsim
0.6\ell_i$.

Figure~\ref{f1} demonstrates that the long-range asymptote of the potential is
dominated by the combined effect of collisions and absorption. It exhibits Coulomb-like
decay $\phi (r) \sim Q_{\rm eff}/r$, where the effective charge $Q_{\rm eff}$ is
determined by the plasma and particle parameters and {\it increases monotonically} in absolute
magnitude with ion collisionality. At short distances the potential follows the DH form
(\ref{Yukawa}), but the actual particle charge $Q$ shows a {\it non-monotonic} dependence on
$\ell_ik_{\rm D}$. In the WC regime $|Q|$ decreases with increasing collisionality, while
in the SC regime $|Q|$ increases until it reaches a certain maximum value when both ion and electron collection by the particle becomes collision dominated~\cite{KhrapakPoP2008}.
Note that in the WC regime the transition from short-range DH to the long-range
Coulomb-like asymptote can occur through an intermediate $\propto r^{-2}$ decay, whilst in the SC
regime the potential is Coulomb-like practically from the particle surface.

\subsection{Effect of ionization/recombination}

So far it has been assumed that there are no plasma sources and sinks
in the vicinity of the particle except at its surface. Physically, this corresponds
to the situation when the characteristic ionization/recombination length is
considerably larger than the characteristic size of the plasma perturbation by the
charged particle, i.e., compensation of plasma losses to the particle occurs very far
from it. However, in real conditions some plasma production and
loss processes always operate in the vicinity of the particle and therefore, it would be interesting to estimate
the importance of these effects.

For an individual particle this has been done in Refs.~\cite{Filippov2007a,Chaudhuri2008a} using the hydrodynamic approach for the case of highly collisional
plasmas. The effects of plasma production and loss processes are introduced in the ion continuity equation
through the corresponding plasma source and loss terms. Electron impact ionization is usually considered as the main mechanism of
plasma production. Plasma loss can be either due to electron-ion volume recombination \cite{Filippov2007a} (which is relevant to high pressure plasmas) or due to
ambipolar diffusion towards the discharge chamber walls and electrodes \cite{Chaudhuri2008a} (which occurs in
low and moderate-pressure gas discharges). In addition, when many particles are present in the system, plasma losses can be associated with absorption on the particles themselves.
This situation has been recently analyzed in detail in Ref.~\cite{else-kom-vla-phys.rev.e-2009}. Let us consider these three cases separately.

\subsubsection{Losses due to volume recombination}
The plasma source and loss terms added to the ion continuity equation are in this case $\nu_In_e-\nu_Rn_en_i$, where $\nu_I$ and $\nu_R$ are the effective ionization and recombination constants, respectively. In the unperturbed state $\nu_I=\nu_Rn$. The standard linearization procedure yields the electric potential of the form
\bee \label{isotropicpotential}
\phi = (Q_+ /r)\exp(-rk_+) + (Q_-/r)\exp(-rk_-),
\ene
where
\bee\label{screeninglength}
k_\pm^2 = \frac{1}{2}\left(k_D^2 +
\frac{\nu_I}{D_i}\right) \pm \frac{1}{2}\sqrt{\left(k_D^2 +
\frac{\nu_I}{D_i}\right)^2 - \frac{4\nu_Ik_{{\rm D}e}^2}{D_i}}
\ene
and
\bee \label{charge}
Q_{\pm} = \mp\frac{Q\lt[k_{\mp}^2 - {k_{\rm D}}^2
- (eJ_i/QD_i)\rt]}{k_+^2 - k_-^2}.
\ene
The potential is screened exponentially but unlike in
Debye-H\"{u}ckel theory it is described by the superposition of the
two exponentials with different inverse screening lengths $k_+$ and
$k_-$. Both these screening lengths depend on the strength of plasma
production (ionization frequency $\nu_I$). The effective charges $Q_+$ and $Q_-$ also depend on
plasma production strength as well as on the ion flux collected by
the grain. The long range asymptote of the potential is determined
by the smaller screening constant $k_-$ with effective charge $Q_-$.
Depending on the strength of plasma production two limiting cases can be considered: low and high
ionization rate.

In the limit of low ionization rate, $\nu_I/D_i \ll
k_{\rm D}^2$, the screening is dominated by
ionization/recombination effects and the screening length
$\lambda_{\rm De}(\ell_ik_{\rm D})\sqrt{\nu/\nu_I}$ is considerably
larger than the electron Debye length since $(\ell_ik_{\rm D}) \gg
\sqrt{\nu_I/\nu}$ in the considered regime. For distances
$\lambda_{\rm D} \ll r \ll \lambda_{\rm De}(\ell_ik_{\rm
D})\sqrt{\nu/\nu_I}$ the potential behaves as Coulomb-like with the
effective charge $Q_- \simeq -(eJ_i/D_ik_{\rm D}^2)$,  i.e., we
recover the result of the previous (no ionization/recombination) limit [Eq. (\ref{potential2})]. Thus, the distance $\lambda_{\rm De}(\ell_ik_{\rm D})\sqrt{\nu/\nu_I}$ determines the length scale below which plasma production is not important and sets up the upper limit of applicability of the results obtained within
the assumption of no ionization/recombination processes in the
vicinity of the grain [Eq. (\ref{potential2})].

In the opposite limit of high ionization rate, $\nu_I/D_i \gg k_{\rm
D}^2$, the screening length is given by the electron Debye
length $\lambda_{\rm De}$ and is independent of the ionization rate
$\nu_I.$ The effective charge $Q_- \simeq Q - eJ_i/\nu_I$ is
somewhat larger in the absolute magnitude than the actual charge.

\subsubsection{Losses due to ambipolar diffusion}
In this case the plasma source and loss terms are $\nu_In_e-\nu_Ln_i$, where $\nu_L$ is the characteristic loss frequency ($\nu_I=\nu_L$). The potential around the grain is \cite{Chaudhuri2008a},
\bee
\phi =
(Q_1/r)\exp(-k_{\rm eff}r) + Q_2/r,
\label{interaction-ambipolar}
\ene
where $Q_1 = Q[1 - \{\nu_I - (e/Q)J_i\}/D_ik_{\rm
eff}^2]$, $Q_2 = Q - Q_1 = (Q\nu_I - eJ_i)/D_ik_{\rm eff}^2$, and
$k_{\rm eff}^2 = {\it k_{\rm D}}^2 + \nu_I/D_i$. Thus, in the considered case the potential is not completely screened, but has
Coulomb-like long-range asymptote. The effective charge $Q_2$
depends both on the strength of ionization $\nu_I$ and the ion flux
$J_i$ collected by the grain. The analysis of the limits of low and high ionization rates is straightforward.

\subsubsection{Losses due to absorption on other grains}
In this case, the role of ion sink is played by
a continuous, immovable, and uniform ``particle medium''. The ion source and
loss terms are essentially the same as in the previous case, with a minor difference that
the ion loss frequency $\nu_L$ may depend on the particle charge which varies with the ion and electron densities.
The principal difference from the previous case
is that $n_i \not = n_e$ in the unperturbed state
due to the presence of the charged particle medium.
This has been shown to change the Coulomb term $Q_2/r$ in Eq.~(\ref{interaction-ambipolar})
to $(Q_2/r) \cos(k_0r)$ as well as to change $Q_{1,2}$ and $k_{\rm eff}$
\cite{RatynskaiaPoP2006,tsytovich-mor-tho-plasma.phys.rep-2002,kompaneetz-tsy-contrib.plasma.phys-2005}.
The parameter $k_0$ is
\bee
k_0 = \sqrt{\frac{P z \nu_L}{\tau D_i[P+z+(d\nu_L/dz)(z/\nu_L)]}}
\label{attraction-wavenumber}
\ene
where $P=(n_i-n_e)/n_i$ is the Havnes parameter
(see, e.g., Eq. (27) of Ref.~\cite{kompaneetz-tsy-contrib.plasma.phys-2005}).
Equation (\ref{attraction-wavenumber}) is derived for
$\nu_L/D_i \ll k_{\rm D}^2$, $\tau \gg 1$, $z \sim 1$, $(d\nu_L/dz)(z/\nu_L) \sim 1$
and yields $k_0 \sim 10^{-2} k_{\rm D}$
for typical experimental parameters
(see also Fig.~3 of Ref.~\cite{RatynskaiaPoP2006}).

However, it has been recently shown that this cosine-like potential
cannot be observed in principle \cite{else-kom-vla-phys.rev.e-2009}.
The reason is that such an ionization-absorption balanced plasma with
$n_i \not = n_e$
is unstable with respect to ion perturbations
and the threshold wavenumber is {\it exactly} equal to $k_0$,
whereas the attraction was derived by implicitly assuming the ion component to be in
a stable equilibrium and considering the screening of a test particle as a static
perturbation of this state.

This instability disappears for a constant ionization source \cite{else-kom-vla-phys.rev.e-2009}.
However, for a constant ionization source
the cosine in the potential changes to the exponent \cite{castaldo-ang-tsy-phys.rev.lett-2006, tsytovich-kom-ang-contrib.plasma.phys-2006}.

\section{Anisotropic plasmas}\label{aniso}

Electric fields are often present in plasmas
(e.g., in rf sheaths, positive column of dc discharge, dc discharge striations, ambipolar electric field in plasma bulk, etc.). This induces an ion drift
and, hence, creates a perturbed region of plasma density downstream from the particle -- the so-called
``plasma wake''. One can apply the linear
dielectric response formalism~\cite{Alexandrov} to calculate the potential distribution in the wake. This
approach is applicable provided ions are weakly coupled to the
particle (i.e. the region of nonlinear ion-particle electric
interaction is small compared to the plasma screening
length). Note that higher ion drift
velocities imply better applicability of the linear theory. The
electrostatic potential created by a point-like charge at rest is
defined in this approximation as
\begin{equation} \label{potential_LR}
\phi ({\bf r})=\frac{Q}{2\pi^2}\int\frac{{\rm e}^{i{\bf kr}}d{\bf
k}}{k^2\varepsilon(0,{\bf k})},
\end{equation}
where $\varepsilon(\omega,{\bf k})$ is the plasma dielectric function. Using a certain model for the dielectric function, one can calculate the
anisotropic potential distribution using analytical approach~\cite{Nambu1995,Vladimirov-Nambu1995,
Vladimirov-Ishihara1996,Ishihara-Vladimirov1997,Xie1999,Lemons2000,Lapenta2000}.
The potential profile can be also obtained from numerical simulations~\cite{Melandso-Goree1995,Lampe2000,Maiorov2001,Winske2001,Lapenta2002,Vladimirov2003,Miloch2008PRE}.
In general, the shape of the wake
potential is sensitive to the ion flow velocity, plasma absorption on the particle,  ion-neutral
collisions, etc. Let us illustrate how the wake potential can depend on plasma conditions using few recent examples from analytic calculations.

The examples below deal with a homogeneous plasma with ion flow driven by an electric field, where
the unperturbed velocity distribution
of ions is determined by the balance of the electric field and collisions with neutrals (mobility-limited flow).
Here, ``homogeneous plasma'' presumes that the inhomogeneity length (e.g. due to Boltzmann distribution of electrons in the field driving the flow) is large enough.
Such a plasma has been shown to be stable with respect to the formation of ion plasma waves,
except for the parameter
range where the thermal Mach number $M_T = u/v_{Tn}$  is larger than $\simeq 8$ and the ratio of the ion-neutral collision frequency to the ion plasma frequency, $\nu/\omega_{pi}$ is less than $\simeq 0.2$ \cite{kompaneets-ivl-vla}. Here $u$
is the ion flow velocity,
$v_{Tn}=\sqrt{T_n/m_i}$ is the thermal velocity of neutrals, $T_n$ is the neutral temperature, and the ion plasma frequency is $\omega_{pi}=\sqrt{4\pi e^2 n/m_i}$.
This justifies the use of the linear
response formalism (\ref{potential_LR}) outside the aforementioned instability range. However, note that the
stability analysis was performed using the BGK collision term,
whereas the use of the more realistic constant-mean-free-path collision term
\cite{else-kom-vla-phys.plasmas-2009} may yield somewhat different instability thresholds.

\subsection{Subthermal ion drifts}

In this subsection we focus on the subthermal ion drift regime, $M_T \lesssim 1$.

The screening in the ``almost collisionless'' case has been investigated using
the kinetic approach with the BGK collision term \cite{kompaneets-mor-ivl-phys.plasmas-2009}.
The ``almost collisionless case'' presumes that the ion-neutral
collision length is much larger than the Debye length and distances where we
want to find the potential. This case is treated by taking the limit $\nu \to 0$ at a fixed $M_T$.
In this limit the role of collisions is only to determine the non-Maxwellian
form of the unperturbed velocity distributions of ions.
The resulting potential for $r \gg \lambda_{\rm D}$
is \cite{kompaneets-mor-ivl-phys.plasmas-2009}
\begin{multline}
\phi({\bf r})=Q\left[\frac{\exp(-r/\lambda_{\rm
D})}{r}-2\sqrt{\frac2{\pi}}\frac{M_T\lambda_{\rm D}^2}{r^3}\cos\theta \right. \\
\left.
+\left(2-\frac{\pi}2\right)\frac{M_T^2\lambda_{\rm
D}^2}{r^3}(1-3\cos^2\theta)\right],\label{222}
\end{multline}
where $\theta$ is the angle between ${\bf r}$ and the ion flow, and $\lambda_D=\sqrt{T_n/(4\pi n e^2)}$ (note a difference with respect to previous notation); the electron component is treated
as a homogeneous neutralizing background which is not perturbed by the presence of the particle.
Equation (\ref{222}) is accurate to $O(M_T/r^5)+O(M_T^3/r^3)$ at $M_T \to 0$, $r \to \infty$.
This result shows:
\begin{itemize}
\item At sufficiently large distances the potential has the
$r^{-3}$-dependence,
which is in agreement with the inverse third power law of screening
in anisotropic collisionless plasmas
\cite{Montgomery1968}.
\item The resulting electric interaction between the particles is non-reciprocal ({\it actio $\not =$ reactio}) since
$\phi({\bf r}) \not = \phi(-{\bf r})$. For instance, if two grains are aligned along the flow
then the grain located downstream may experience attraction whereas the grain located upstream
will be always repelled.
\item The potential in the direction perpendicular to the flow does not have an attractive part.
This is contrast to the case of a shifted Maxwellian distribution (see, e.g., Eq.~(19) of Ref.~\cite{cooper-phys.fluids-1969}, Fig.~2 of
Ref.~\cite{kompaneets-vla-ivl-new.j.phys-2008}, or Fig.~3 of Ref.~\cite{wang-joy-nic-j.plasma.phys-1981}) and shows the importance of
accounting for the non-Maxwellian form of the ion distribution.
\end{itemize}

\begin{figure}[!t]
\centering
\includegraphics[width = \linewidth]{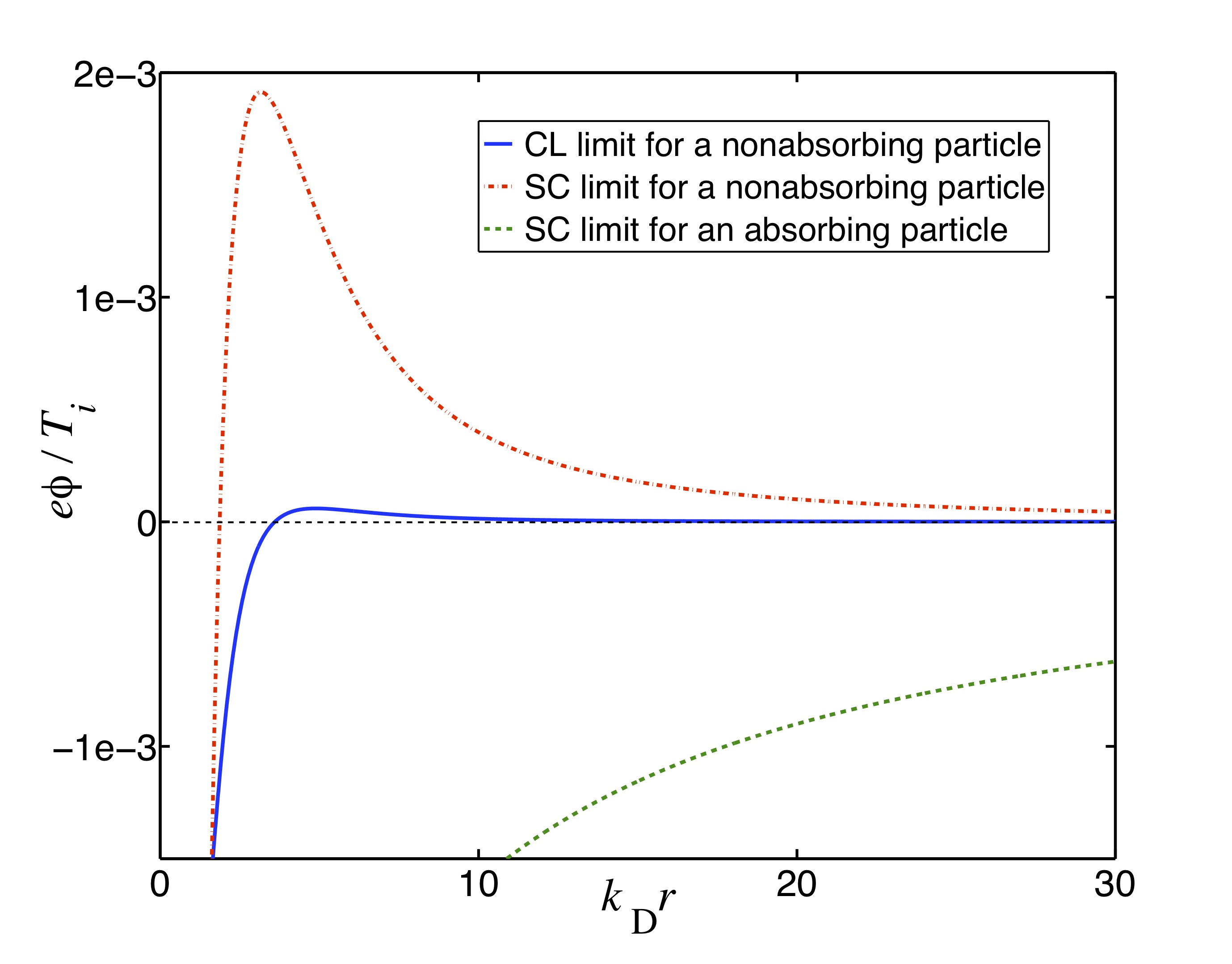}
\caption{Distribution of the normalized electric potential behind a small individual charged particle in a weakly anisotropic plasma. The solid curve represents the potential distribution in the collisionless (CL) limit, Eq.~(\ref{222}), for $|Q|e/(aT_i) = 2$ (here $T_i \equiv T_n$), $M_T = 0.2$, and $r > 1.2 \lambda_{\rm D}$. The dash-dotted curve corresponds to the strongly collisional (SC) limit for a nonabsorbing particle. The dashed curve shows the potential for an absorbing particle in the same highly collisional limit. Both last curves are calculated using Eq.~(\ref{normal}) for the same parameters as for the CL case but with $T_e = T_i$, $a/\lambda_{\rm D} = 0.02$, $\ell_i/\lambda_{\rm D} = 0.1$.}\label{f2}
\end{figure}

The wake effect has been proposed to be used to design binary interactions in complex plasmas
by applying electric fields of various polarizations \cite{kompaneets-mor-ivl-phys.plasmas-2009}.
The idea is to apply an electric field oscillating with a frequency which is (i) much faster than
that characterizing the particle dynamics and (ii) much
slower than the ion plasma frequency and the ion-neutral collision
frequency. In this case the particles do not react to the field, whereas
ions react instantaneously and the
resulting interaction between the particles is determined
by the time-averaged distribution of the electric potential in the
screening clouds. Here the most interesting case is the case of spherical polarization
where the vector of the electric field rotates in such a way that its directions are
isotropically distributed in 3D space but its absolute value remains constant.
In this case the resulting potential is isotropic but the second and third terms
in Eq.~(\ref{222}) are averaged out
so that the effect of the field is in the rest term $O(M_T/r^5)+O(M_T^3/r^3)$. The resulting potential
was investigated numerically in Ref.~\cite{kompaneets-mor-ivl-phys.plasmas-2009}
for finite $M_T$ and found to have an attractive part.

The effect of ion absorption on the particle surface has been so far investigated only
for subthermal ion flows in highly collisional plasmas using the hydrodynamic approach~\cite{Chaudhuri2007}. In this regime the absorption-induced ion rarefication behind the particle can overcome the effect of ion focusing and a negative space charge region develops downstream from
the particle. This can have important consequences for particle motion in low-ionized plasmas \cite{VladimirovPRL2007} and, therefore, let us discuss this issue in some more detail.

A stationary negatively charged spherical point-like particle which is immersed in a
quasineutral highly collisional plasma is considered. The electric field is sufficiently weak so that the ions are drifting with subthermal velocity while electrons form stationary background. Plasma absorption occurs on the grain
surface, i.e. it acts as a plasma sink. Ionization/recombination processes in the vicinity of the grain are neglected. The potential distribution is obtained by solving Poisson equation coupled to the hydrodynamic equations for ions and Boltzmann equation for electrons using the standard linearization technique under the assumption $M_T\ll k_{\rm D}\ell_i\ll 1$. Further, using the known asymptotic expression for the ion flux to an infinitesimally small grain
($a\ll\lambda_{\rm D})$ in the continuum limit ($\ell_i\ll a$) \cite{Su-Lam1963,Khrapak2006b,Chang} the expression for the potential downstream from the particle ($\theta=0$) can be written in the form \cite{Chaudhuri2007}:
\begin{multline}\label{normal}
\phi({\bf r}) = Q\left[\frac{\exp(-rk_{\rm D})}{r}  \right. \\ - \left(\frac{M_Tk_{{\rm D}i}^2}{2\ell_ik_{\rm D}^4}\right)\frac{2 - (r^2k_{\rm D}^2 + 2rk_{\rm D} + 2)\exp(-rk_{\rm D})}{r^2} \\+\lt(\frac{k_{{\rm D}i}}{k_{\rm D}}\rt)^2\frac{1 - \exp(-rk_{\rm D})}{r} + \lt(1 - \frac{2k_{\rm De}^2}{k_{\rm D}^2}\rt)\left(\frac{M_Tk_{{\rm D}i}^2}{2\ell_ik_{\rm D}^4}\right)\\ \left. \times\frac{2 - \lt[(1 - k_{\rm De}^2/k_{\rm Di}^2)^{-1}r^2k_{\rm D}^2 + 2rk_{\rm D}+2\rt]\exp(-rk_{\rm D})}{r^2}\right]
\end{multline}
Let us briefly analyze the structure of the Eq.~(\ref{normal}). The first term is the usual isotropic
Debye-H\"{u}ckel potential and the second term represents the anisotropic part of the potential
behind a nonabsorbing particle. The third and fourth terms represent, respectively, the isotropic and
anisotropic parts of the electric potential associated with the effect of ion absorption.
In the absence of ion absorption the long-range asymptote of the electric
potential can be written in the form
\begin{equation}\label{Stenflo}
\phi \approx - (Qu\nu/\omega_{pi}^2r^2)(k_{{\rm D}i}/k_{\rm D})^4.
\end{equation}
Except of the factor $(k_{{\rm D}i}/k_{\rm D})^4$ this expression is identical to that obtained in Ref.~\cite{Stenflo} for the potential behind a slowly moving non-absorbing test charge
in collisional plasma. The difference appears because in Ref.~\cite{Stenflo} only electron screening was taken into account. Thus, in contrast to the collisionless regime, the potential of a non-absorbing particle in collisional plasmas exhibits $\propto r^{-2}$ long-range decay. In the presence of absorption the long-range asymptote of the
anisotropic part of the potential can be written as
\begin{displaymath}
\phi \approx -
2(Qu\nu/\omega_{pi}^2r^2)(k_{{\rm D}i}/k_{\rm D})^4(k_{{\rm D}e}/k_{\rm D})^2.
\end{displaymath}
Thus, the effect of absorption induced ion rarefication competes with the effect of ion focusing and the amplitude of the anisotropic part of potential decreases, provided $T_e>T_i$.  More important is the contribution from the isotropic part associated with absorption. It is always negative (for a negatively charged particle) and can be written as
\begin{displaymath}
\phi\approx  (Q/r)(k_{{\rm D}i}/k_{\rm D})^2,
\end{displaymath}
This contribution is most often dominant and completely determines the long-range behavior of the potential~\cite{Chaudhuri2007}.

Figure \ref{f2} shows an example of calculating the electric potential downstream from the particle for the situations discussed above. The solid curve corresponds to the non-absorbing particle in  collisionless plasma [Eq. (\ref{222})]. Ions are focused downstream from the particle and a positive space charge region exists. The dash-dotted curve shows the potential distribution for a non-absorbing particle in highly collisional plasmas [first two terms of Eq. (\ref{normal})]. Here collisions enhance ion focusing \cite{IvlevPRL} and the positive space charge region grows considerably. However, when absorption is included using Eq. (\ref{normal}), the absorption induced ion rarefication plays a dominant role and the potential is negative (dashed curve), in contrast to the previous cases.

\subsection{Suprathermal ion drifts}

\begin{figure}[!t]
\centering
\includegraphics[width = 0.8\linewidth]{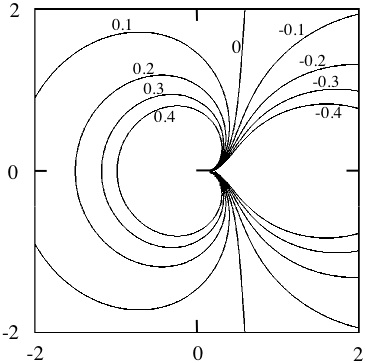}
\caption{Contour plot of the potential around a small charged dust particle in a weakly ionized plasma with suprathermal ion drift driven by an electric field [Eq. (\ref{potential-sigma-const})] for $\lambda_{\rm D, eff}/\ell_i = 0.5$.  The charge is in the center and the drift direction is from the left to the right.  The distance and the potential are in units of $\lambda_{\rm D, eff}$ and $Q/\lambda_{\rm D, eff}$, respectively.}\label{f3}
\end{figure}

In the highly suprathermal regime $M_T\gg1$
the potential can be found analytically using the kinetic approach
with the realistic constant-mean-free-path collision term \cite{kompaneets-kon-ivl-phys.plasmas-2007}:
\begin{multline}
\phi({\bf r})= \frac{2Q}{\pi\ell_i}{\rm Re}\int_{0}^{\infty} \, dt \frac{\exp[it(r_\|/\ell_i)]}{
1+(\ell_i/\lambda_{\rm D,eff})^2Y(t)} \\
\times K_0 \left(\frac{r_\perp}{\ell_i}\sqrt{\frac{t^2+(\ell_i/\lambda_{\rm D,eff})^2X(t)}{1+(\ell_i/\lambda_{\rm D,eff})^2Y(t)} } \right).
\label{potential-sigma-const}
\end{multline}
Here $r_\perp$ is the distance in the plane perpendicular to the flow,
$r_\|$ is the distance along the flow ($r_\|>0$ and $r_\|<0$ along and against the flow, respectively),
$\lambda_{\rm D, eff}=[eE\ell_i/(4\pi n e^2)]^{1/2}$ is the effective Debye length,
$E$ is the field which drives the flow [it is related to the flow velocity
via $u=\sqrt{2eE\ell_i/(\pi m_i)}$],
$K_0$ is the zero-order modified Bessel function of the second kind
\cite{abramowitz-stegun-1972},
\begin{multline}
X(t)=1-\sqrt{1+it}, \\
{\displaystyle Y(t)=\frac{2\sqrt{1+it}}{it}\int_0^1\frac{d\alpha}{[1+it(1-\alpha^2)]^2}-\frac{1}{it(1+it)},}\label{xyi}
\end{multline}
and the square roots must be taken with positive real part; the electrons are considered
as a homogeneous background which is not perturbed by the presence of particle.
Expression (\ref{potential-sigma-const}) works at all but very small angles with respect to the flow
since it diverges logarithmically at $r_\perp \to 0$ for $r_\|>0$ due to the neglect of the thermal spread of neutral velocities. The contor plot of potential (\ref{potential-sigma-const}) is shown in Fig.~\ref{f3}.

The derived potential applies to the screening of charged particles in plasma presheaths where the ion flow velocity is in between
the thermal velocity of neutrals and the Bohm velocity. This potential has been shown in
Ref.~\cite{kompaneets-kon-ivl-phys.plasmas-2007} to be in agreement with measurements of
Konopka et al. \cite{konopka-mor-rat-phys.rev.lett-2000}.

The potential (\ref{potential-sigma-const}) at large distances
is \cite{kompaneets-kon-ivl-phys.plasmas-2007}
\begin{equation}
\label{farfield}
\phi({\bf r})= -\frac{Q\lambda_{\rm D,eff}^2}{\ell_i}\frac{\cos\theta}{r^2}\left(\frac2{1+\cos^2\theta}\right)^{3/2}
\end{equation}
which is accurate to $O\left(r^{-3}\right)$ at $r \to \infty$.
This is an unscreened dipole-like field with the dipole moment
$|Q|\lambda_{\rm D,eff}^2/\ell_i$ directed along the flow for $Q<0$, although
there is a difference from a pure dipole field due to the factor $[2/(1+\cos^2\theta)]^{3/2}$.
The $\propto r^{-2}$-dependence in Eq.~(\ref{farfield}) is again different from the inverse third power law decay in
collisionless anisotropic plasmas because of a finite collision length.
Note that for $\theta =0$ Eq. (\ref{farfield}) is identical to the result of hydrodynamic approach [Eq. (\ref{Stenflo})], provided we assume $T_e\rightarrow \infty$ and $u=eE/m_i\nu$ in Eq. (\ref{Stenflo}). In both cases the dipole momentum is $QeE/(m_i\omega_{pi}^2)$.

\section{Summary}\label{concl}

In this paper, we have summarized and discussed our recent results regarding the electric potential distribution around a small charged particle
in weakly ionized plasmas. Different effects influencing the shape of the potential, including plasma absorption on the particle, ion-neutral collisions, plasma production and loss processes, and plasma anisotropy have been considered. The generic property of the results obtained so far is that sufficiently close to the particle the potential can be well approximated by the Debye-H\"{u}ckel (Yukawa) form. At longer distances, the potential usually exhibits power-law decay $\propto c/r^n$. The value of $n$ ($n=1, 2$, or $3$ in the cases investigated), as well as the sign and magnitude of the parameter $c$, depend on the plasma and particle properties. These results have important consequences for plasma-particle and interparticle interactions and related phenomena, including phase transitions, phase diagrams, transport, waves, etc.





%

%

\begin{IEEEbiography}[{\includegraphics[width=1in,height=1.25in,clip,keepaspectratio]{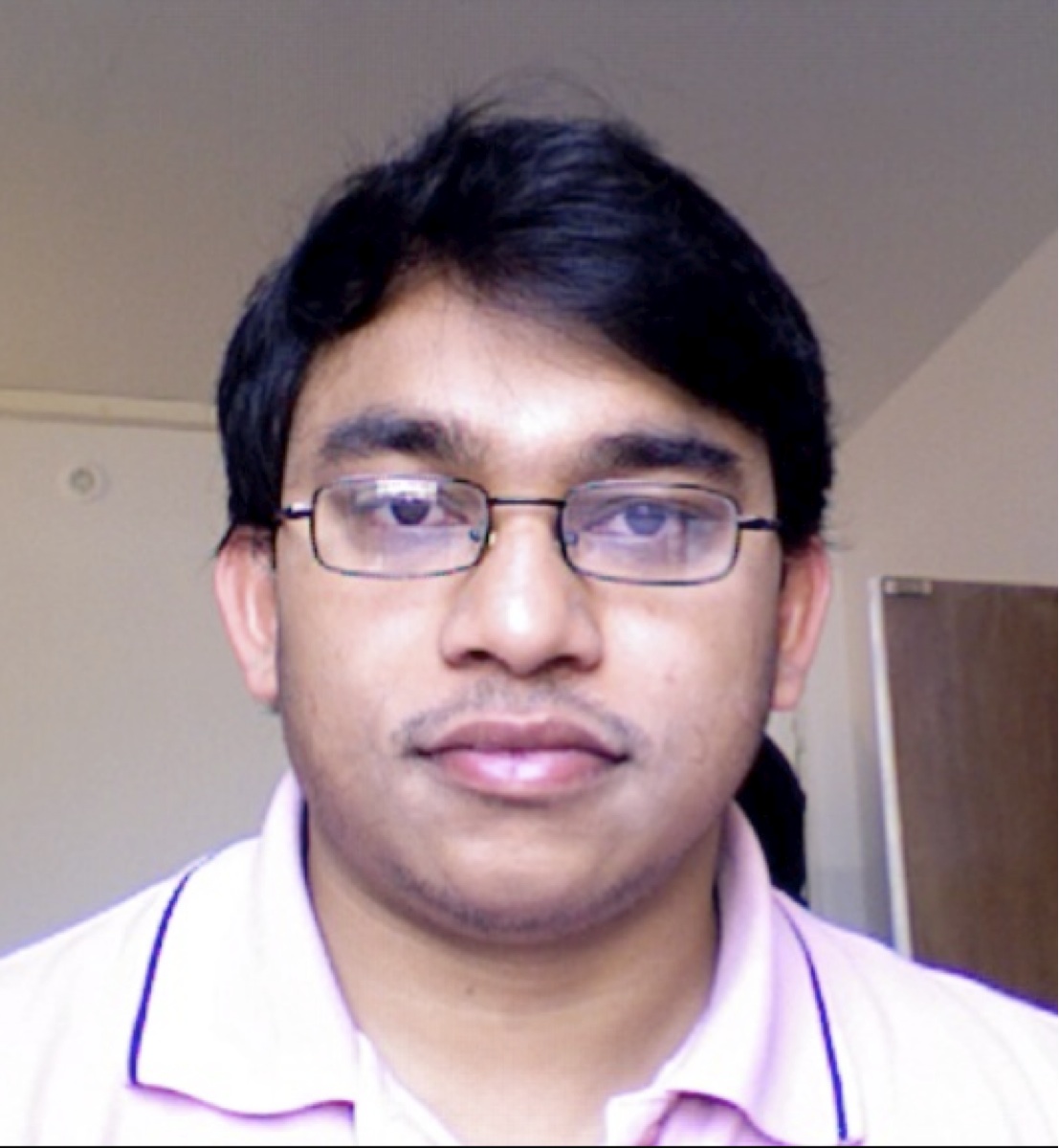}}]
{\bf Manis Chaudhuri} was born in Hooghly, India in 1978. He received the M.Sc. degree in physics
from the University of Calcutta, India in 2001 and the Ph.D. degree in
physics from the Ludwig-Maximilians-University, Munich, Germany in 2008.

From 2003 to 2005, he was a Research Fellow at Saha Institute of Nuclear Physics, Calcutta, India.
Since 2005 he has been with the theory (complex plasma) group at Max-Planck-Institute for Extraterrestrial Physics, Garching, Germany.
His research interest includes theoretical and experimental investigations of complex plasmas.
\end{IEEEbiography}

\begin{IEEEbiography}[{\includegraphics[width=1in,height=1.25in,clip,keepaspectratio]{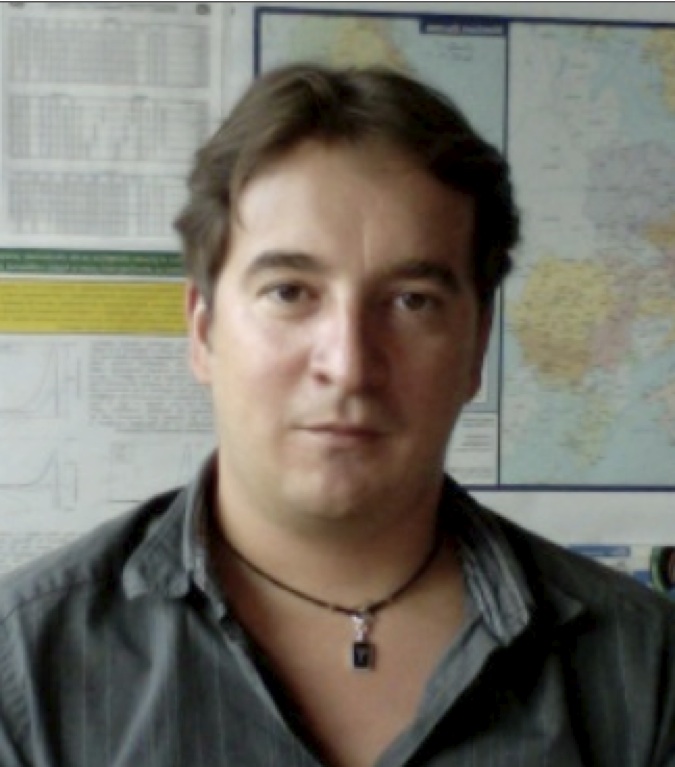}}]
{\bf Sergey Khrapak} was born in Moscow, Russia, in 1973. He received the M. Sc. and Ph.D. degrees in physics from the Moscow Institute of Physics and Technology, Moscow, in 1996 and 1999, respectively.

From 1996 to 2000, he was a Research Fellow with High Energy Density Research Center, Institute for High Temperatures, Russian Academy of Sciences, Moscow. Since 2000, he has been with the theory (complex plasma) group of Max-Planck-Institut f\"ur extraterrestrische Physik, Garching, Germany. He is the author of more than 100 scientific papers and more than 60 contributions to international and national conferences. His scientific interests are mostly focused on the theory of complex (dusty) plasmas.

\end{IEEEbiography}

\begin{IEEEbiography}[{\includegraphics[width=1in,height=1.25in,clip,keepaspectratio]{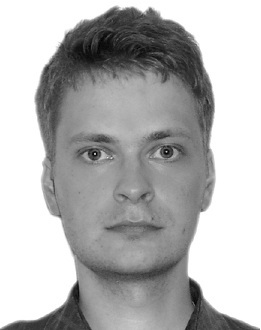}}]
{\bf Roman Kompaneets} was born in 1980 in Moscow, U.S.S.R. He received
the MS degree in applied mathematics and physics from the Moscow Institute of Physics and Technology, Russia,
in 2003
and
the Dr. rer. nat. degree (PhD equivalent) in physics from the Ludwig Maximilians University of Munich, Germany, in 2007.
He is a Professor
Harry Messel Research Fellow in the School of Physics at the University of Sydney, Australia.
His research interests are the
collective kinetic effects in various plasmas including dusty plasmas.

\end{IEEEbiography}

\begin{IEEEbiography}[{\includegraphics[width=1in,height=1.25in,clip,keepaspectratio]{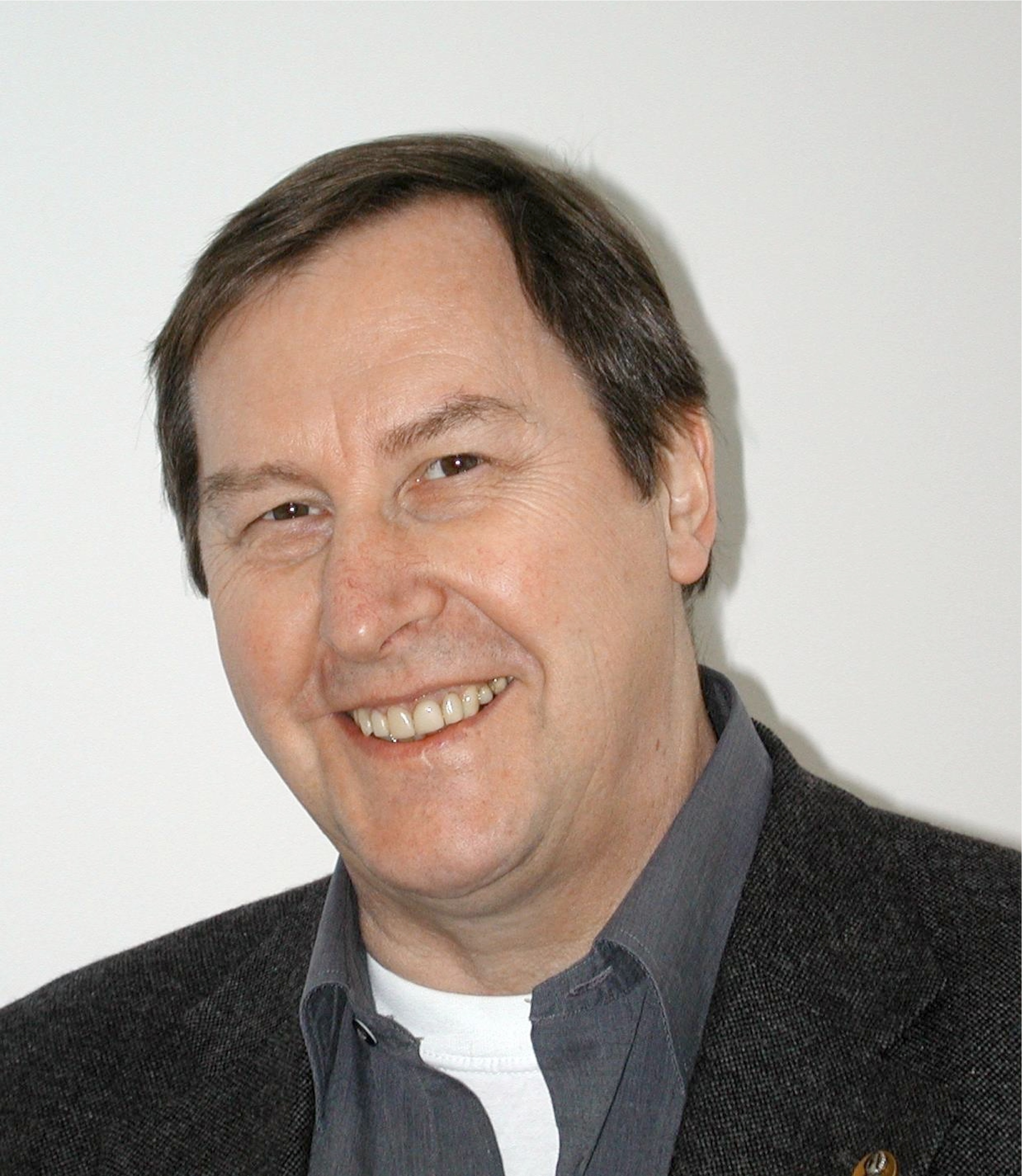}}]
{\bf Gregor Morfill} was born in Oberhausen, Germany, in 1945. He received the B.
Sc. degree in physics and the Ph.D. degree in space plasma physics from Imperial College, London University, London, U.K.,
in 1967 and 1970, respectively, the Honorary Doctorate degree from the Technical University of Berlin, Berlin, Germany, in 2003.

Since 1984, he has been the Director of Max-Planck Institut f\"ur extraterrestrische Physik, Garching, Germany. He also holds honorary Professorships at the University of Leeds, Leeds, U.K., the University of Arizona, Tucson. He is the author of more than 500 scientific papers. His present scientific interests are mostly focused on complex (dusty) plasmas, astrophysical plasmas, and
plasma applications in medicine.

Professor Morfill is a foreign member of the Russian
Academy of Sciences. He has the recipient of a number of important prizes including Patten
Prize, Bavarian Innovation Prize, Wissenschaftpreis of the German ``Stifterverband", Gagarin Medal, Ziolkowski Medal, etc.
\end{IEEEbiography}






\end{document}